# Pre-Breakdown Processes in Dielectric Fluid in Inhomogeneous Pulsed Electric Fields

Mikhail N. Shneider[1,*] and Mikhail Pekker[2]

[1] *Department of Mechanical and Aerospace Engineering, Princeton University, Princeton, NJ, 08544*
[2] *MMSolution, 6808 Walker Street, Philadelphia, PA, 19135*

**We consider the development of pre-breakdown cavitation nanopores appearing in the dielectric fluid under the influence of the electrostrictive stresses in the inhomogeneous pulsed electric field. It is shown that three characteristic regions can be distinguished near the needle electrode. In the first region, where the electric field gradient is greatest, the cavitation nanopores, occurring during the voltage nanosecond pulse, may grow to the size at which an electron accelerated by the field inside the pores can acquire enough energy for excitation and ionization of the liquid on the opposite pore wall, i.e., the breakdown conditions are satisfied. In the second region, the negative pressure caused by the electrostriction is large enough for the cavitation initiation (which can be registered by optical methods), but, during the voltage pulse, the pores do not reach the size at which the potential difference across their borders becomes sufficient for ionization or excitation of water molecules. And, in the third, the development of cavitation is impossible, due to an insufficient level of the negative pressure: in this area, the spontaneously occurring micropores do not grow and collapse under the influence of surface tension forces. This paper discusses the expansion dynamics of the cavitation pores and their most probable shape.**

## I. Introduction.

A mechanism for the rapid breakdown in the fluid, associated with the occurrence of cavitation ruptures under the influence of the electrostrictive forces near the needle electrode, was proposed in [1]. Later, a hydrodynamic model of compressible fluid motion under the influence of the ponderomotive electrostrictive forces in a non-uniform time dependent electric field was suggested in [2]. As shown in [2], if the voltage on the needle electrode is growing fast enough (a few nanoseconds), the negative stress is created in liquid, and it may be sufficient for the cavitation formation. A nanosecond breakdown, the beginning of which can be explained by the formation of cavitation in a stretched liquid due to electrostriction, was investigated experimentally in [3-7]. In [8], based on a theoretical model [2], it had been shown experimentally that the initial stage of development of a nanosecond breakdown in liquids is associated with the appearance of discontinuities in the liquid (cavitation) under the influence of electrostriction forces. The comparison of the experimentally measured area dimensions and its temporal development was found to be in a good agreement with the theoretical calculations. A theory of the cavitation initiation in inhomogeneous pulsed electric field was developed in [9], as well as the method that allows to determine the critical parameters, at which cavitation begins, on the basis of the comparison between the experiment and the simulation results within the framework of hydrodynamics of compressible fluids, was proposed.

In Part II, it is shown in the model example of a spherical electrode that if the applied voltage on the electrode is growing fast enough (~ few nanoseconds), the negative stress induced in liquid by the ponderomotive electrostrictive forces does not have time to be compensated by the rising of the positive hydrostatic pressure, related to the fluid influx into the negative pressure region. Therefore, the conditions arise for the formation of the fluid discontinuities (nanovoids), i.e., cavitation.

* m.n.shneider@gmail.com

Part III presents the computed probabilities of the critical bubble appearance in the vicinity of the electrode at different time moments

Part IV presents the equations for the nanovoid expansion in the presence of a strong inhomogeneous electric field and the approaches to this problem, which will be considered, are formulated. The presented numerical results of the nanopores expansion in the case of the spherical electrode can be easily generalized to an arbitrarily shaped electrode. It is shown that there are three characteristic regions in the vicinity of the electrode. In the first region, where the electric field gradient is greatest, the occurring cavitation nanopores have enough time during the nanosecond voltage pulse to grow to a size at which an electron can gain enough energy for the excitation and ionization of the liquid molecules on the pore wall. In the second region, the electrostrictive negative pressure reaches values at which the cavitation development becomes possible (that can be registered by the optical methods), but the nanovoids, appearing during the voltage pulse, do not have enough time to grow to the size at which the potential difference across their borders becomes sufficient for the ionization or excitation of water molecules. And, in the third, the development of cavitation is impossible, since the spontaneously occurring nanovoids do not grow, because the value of the electrostrictive negative pressure is relatively small and cannot compete with the forces of surface tension.

Part V discusses the form of an expanding cavitation micropore. It is shown that a perturbation of the electric field caused by the micropores near the equator (in the plane perpendicular to the direction of the undisturbed electric field) produces a flow directed to the pore, and near the poles – out of the pore. The velocities of the arising flow are comparable to the velocities of the spherical pores expansion and an order of magnitude greater than the flow velocities caused by the unperturbed nonuniform electric field due to the applied voltage on the electrode. As a consequence, the micropore is stretched along the electric field, as its extension is decelerated in the equatorial plane due to the counterflow of the liquid, and the electrostrictive negative pressure is compensated by the hydrostatic pressure.

Note, that the evolution of gas filled bubbles in a dielectric liquid within a strong electric field is theoretically and experimentally studied [10, 11]. Their elongation along the field was due to the fact that the component of the compressive pressure on the poles of the electric field in a spherical bubble in $1/\varepsilon$ ($\varepsilon$ is the dielectric permittivity of the liquid) times smaller than at the equator. In these papers, the effect of the ponderomotive force on the bubble has not been considered because in a stationary (or slowly varying in time) electric field, the negative pressure in the fluid is compensated by the hydrostatic pressure. In this paper, we consider the case of a rapid variation of the nonuniform electric field when the ponderomotive forces cannot be neglected and they (not the gas in the bubble) lead to a rapid expansion of the micropore.

In the final part of this work, the conclusions are formulated.

## II. Motion of the fluid under the action of the ponderomotive forces

The dynamics of a dielectric liquid (water) in a pulsed inhomogeneous electric field in the approximation of the compressible fluid dynamics can be consider within the standard system of the continuity of mass and momentum equations [12]:

$$\frac{\partial \rho}{\partial t} + \nabla(\rho \vec{u}) = 0$$

$$\rho \frac{\partial \vec{u}}{\partial t} + \rho(\vec{u} \cdot \nabla)\vec{u} = -\frac{\varepsilon_0}{2} E^2 \nabla \varepsilon + \frac{\varepsilon_0}{2} \nabla\left( E^2 \frac{\partial \varepsilon}{\partial \rho} \rho \right) - \nabla p + \nu \Delta \vec{u} , \quad (1)$$

where the first and the second terms in the right part of the equation (1) are the volumetric densities of ponderomotive forces [13,14,15], $\varepsilon_0$ is the vacuum dielectric permittivity, $\rho$ is the liquid density, $\vec{u}$ - velocity, $\vec{E}$ is the electric field, $\nu = 10^{-6}$ m$^2$/s is the kinematic viscosity of water. The relation of the hydrodynamic pressure $p$ with the compressible fluid density $\rho$ (in our case - water) is given by the Tait equation of the state [16,17]:

$$p = (p_0 + B)\left( \frac{\rho}{\rho_0} \right)^{\gamma} - B , \quad (2)$$

$\rho_0 = 1000$ kg/m$^3$, $p_0 = 10^5$ Pa, $B = 3.07 \cdot 10^8$ Pa, $\gamma = 7.5$

For the polar dielectrics fluid (water)

$$\frac{\partial \varepsilon}{\partial \rho} \rho = \alpha \varepsilon \quad (3)$$

where $\alpha \sim 1.3 - 1.5$ is the empirical factor for most of the studied polar dielectric liquids, including water, [18, 19] and for the nonpolar dielectrics $\alpha = (\varepsilon - 1) \cdot (\varepsilon + 2)/3\varepsilon$ [20].

In the Figures 1 and 2, the results of the calculations performed for a spherical electrode of radius $r_0 = 100\ \mu$m are shown for the voltage amplitude $U_0 = 54$ kV, $\varepsilon = 81$ and the electric field varying linearly with time:

$$E(r,t) = \frac{U_0 r_0}{r^2} \frac{t}{t_0} , \quad (4)$$

Here, $t_0$ is the duration of the pulse front. The results shown in Fig. 1 correspond to $t_0 = 3ns$, and in Fig. 2, to $t_0 = 100ns$.

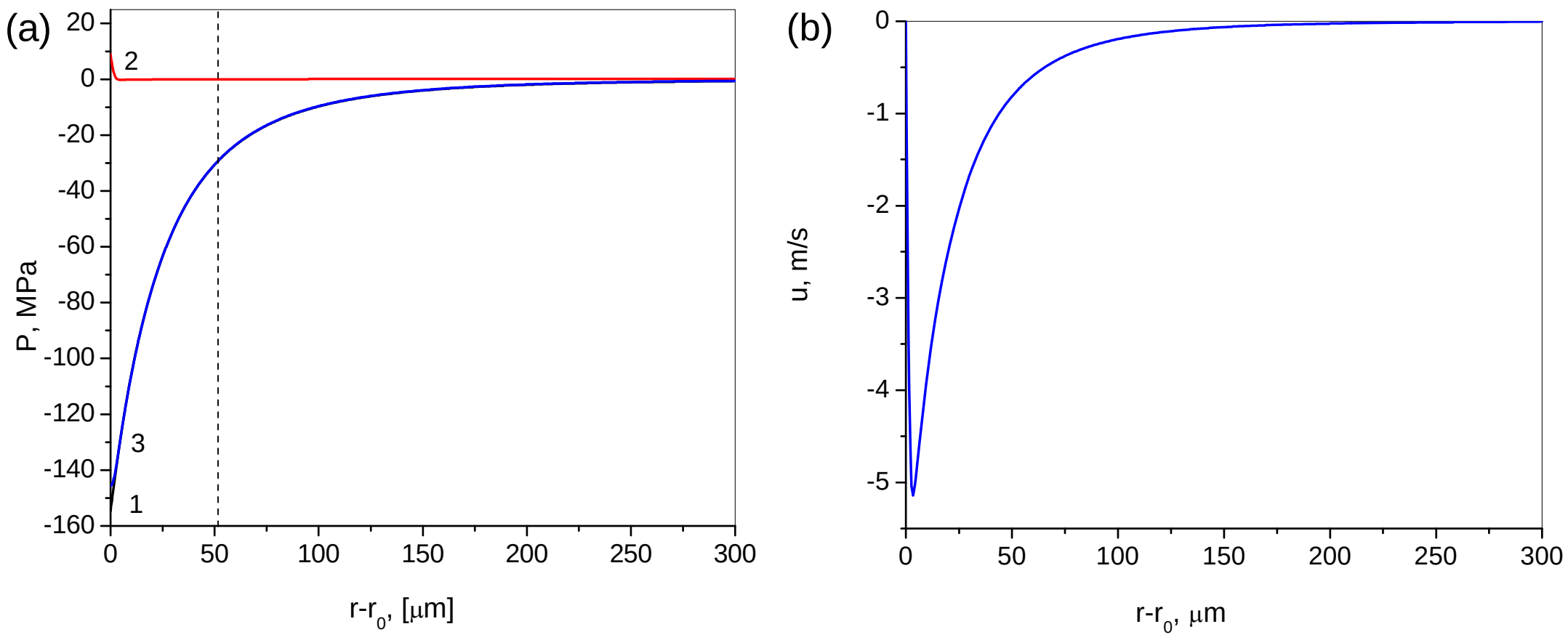

**Fig. 1.** (a) The pressure dependence on the distance from the electrode at $t$=$t_0$=3ns. 1 - is the pressure associated with the electrostriction forces ($P_E = -0.5\alpha\varepsilon_0\varepsilon E^2$), 2 – the hydrostatic pressure associated with the change in density of the fluid, 3 - the total pressure equal to the sum of the electrostrictive and the hydrostatic pressures. The dotted line indicates the distance, at which the pressure is greater than the critical, assumed, for example, $P_{cr} \approx$ -30 MPa. (b) The dependence of the velocity of fluid from the electrode distance.

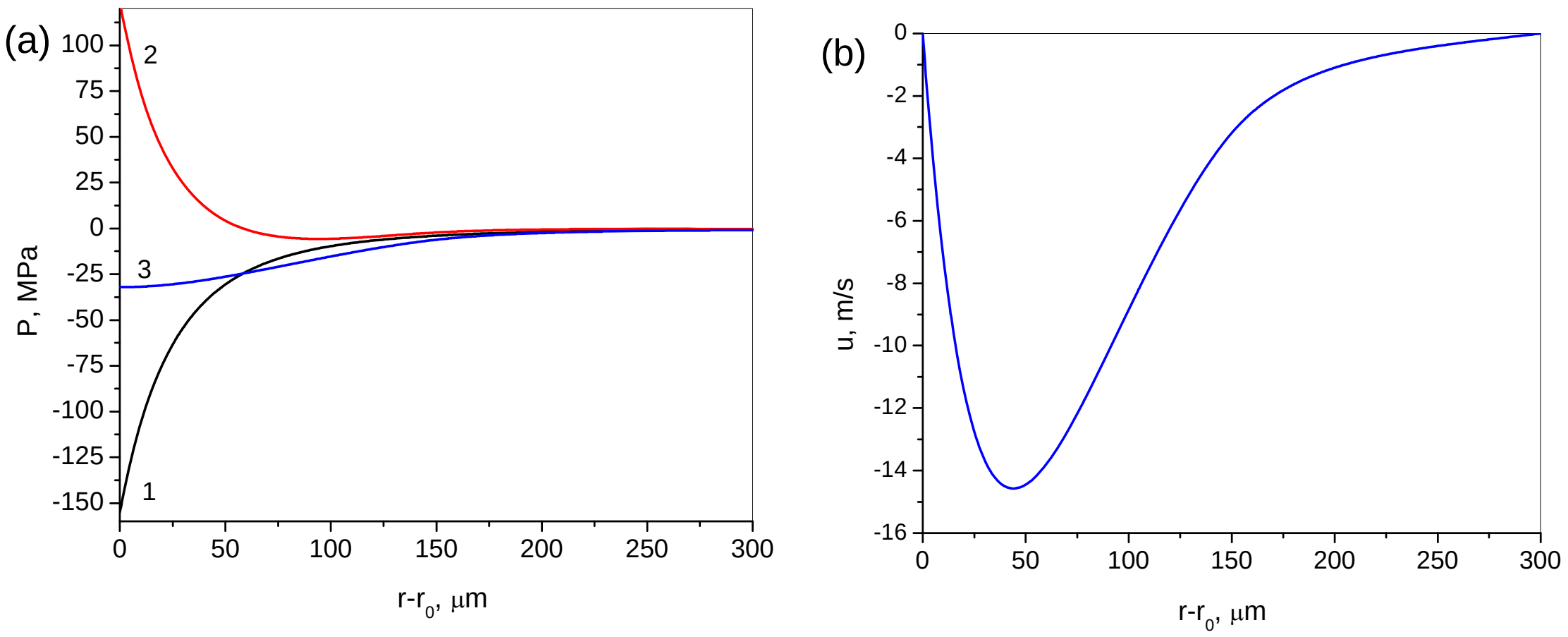

**Fig. 2.** Same as in Fig. 1, at $t$=$t_0$=100ns. There is no region of developed cavitation.

With the rapid rise of the applied voltage on the electrode (on the order of a few nanoseconds), the liquid (water) does not have enough time to come in motion and results in a change of the hydrostatic pressure, compensating the ponderomotive force (Fig. 1). Consequently, the ponderomotive forces cause a significant electrostrictive tensile stress in the dielectric liquid, which can lead to a disruption of the continuity of liquid (creating nanopores), similar to those observed in [21, 22, 23]. For the formation of discontinuities, it is necessary that the absolute value of the negative pressure in the fluid reach the values on the order of 10–30MPa [22]. In this case, the density fluctuations, which always exist due to thermal motion, lead to the formation of growing cavitation voids.

With a relatively slow increase in the voltage on the electrode, the fluid moves to the electrode, and the total pressure does not reach the critical value, at which the conditions required to initiate the fluid discontinuities (Fig. 2) are created. It should be noted that with the rapid and slow increase in voltage, the fluid velocity is less than the speed of sound by more than two orders of magnitude (Fig. 1 (b), 2 (b)) and less than the expansion velocity of nanopores by more than an order of magnitude (see Part III).

If the voltage on the electrodes grows relatively slowly (the voltage pulse front durations are hundreds nanoseconds or longer), the influence of ponderomotive forces on the fluid dynamics can be neglected. In this case, the conditions of works [10,11], in which the dynamics of a gas bubble in a strong electric field was studied, are valid.

## III. The probability of the critical bubble appearance

The probability of the appearance of the critical bubble in the volume $V$ over the time $t$ due to the development of thermal fluctuations, in accordance with the theory [24, 25], is

$$W_{pore} = 1 - \exp\left( -\int_0^t \int_V \Gamma dt_1 dV \right). \tag{5}$$

Here, $\Gamma$ [$m^{-3}s^{-1}$] characterizes the rate of the cavitation voids appearance in unit volume per second.

$$\Gamma = \frac{3k_B T}{16\pi(\sigma\cdot k_\sigma)^3}\frac{|P|^3}{4\pi\hbar}\exp\left(-\frac{16\pi(k_\sigma\sigma)^3}{3k_B T\cdot P^2}\right) \qquad (6)$$

Here, $T$ is the temperature of fluid in Kelvin, $k_B$ is the Boltzmann constant, $\hbar$ - Plank constant. The parameter $k_\sigma = 1/(1+2\delta/R_{cr})$ characterizes the dependence of the coefficient of surface tension on the critical radius of the nanopores $R_{cr}$, which is determined by the balance of the tensile and the surface tension forces [9], $\delta$ is the so-called Tolman coefficient [26], which is determined from the experiment.

If the critical pressure, at which cavitation begins is $P_{cr} \approx -30$ MPa [22], then, as follows from [9], $k_\sigma \approx 0.26$, $R_{cr} \approx 1.24$ nm, and the Tolman coefficient $\delta = 1.8$ nm. That allows calculating the probability of cavitation, the region where cavitation develops, and the concentration of the cavitation ruptures generated during the voltage pulse. Figure 3 shows the normalized dependencies of the rate of generation of cavitation voids $\Gamma$ on the distance from the electrode at different time moments during the voltage pulse. The calculations were made at $t_0 = 3ns$, $r_0 = 100$ μm, and $U_0 = 54$ kV. For all generated cavitation nanopores, regardless of the local negative pressure value $|P(r,t)| > |P_{cr}|$, $k_\sigma = 0.26$ was assumed. Note, that we do not take into account that the rate of cavitation voids generation near the electrode reaches saturation or dramatically decreases, during the voltage pulse, due to the breakdown development.

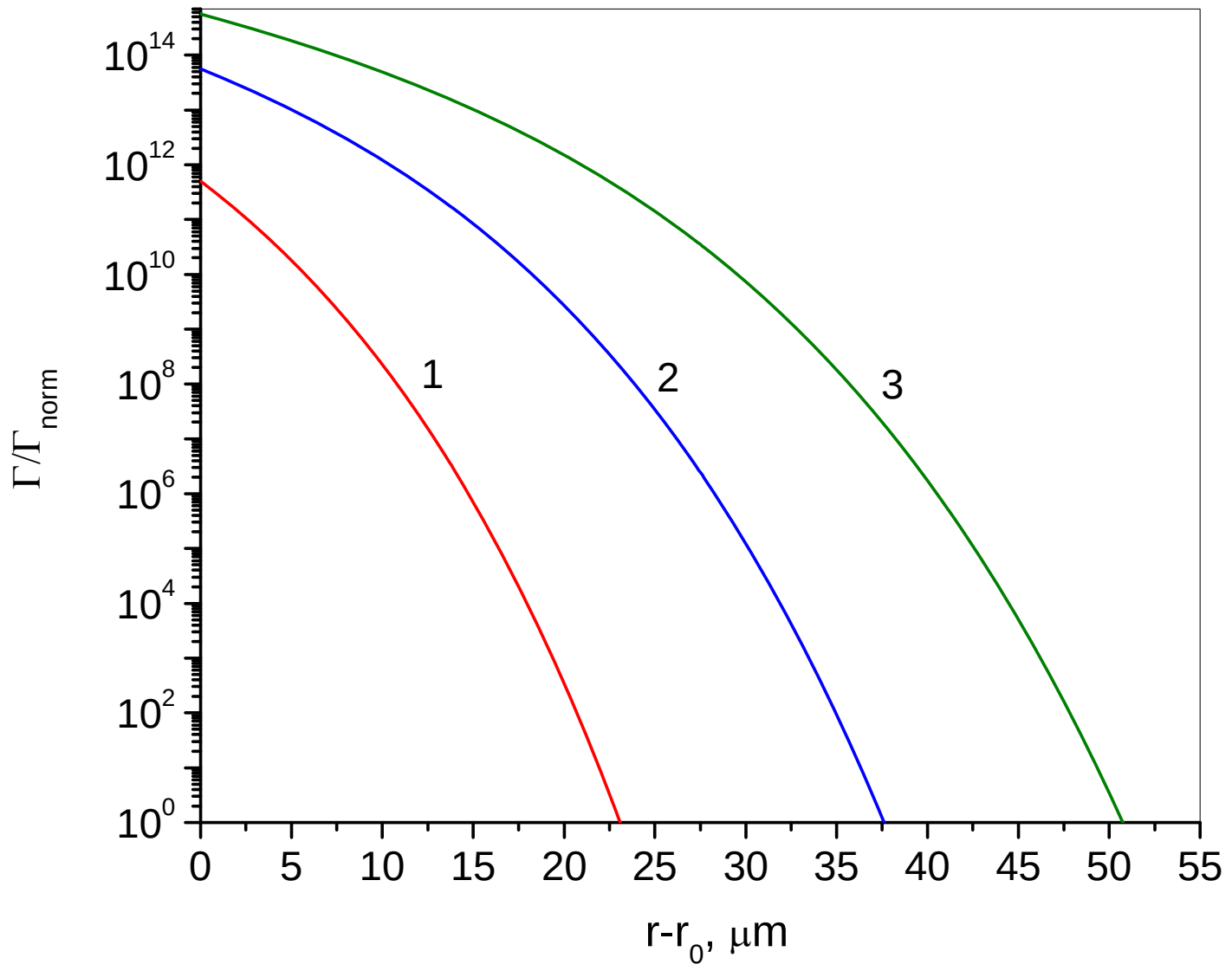

**Fig. 3**. The computed dependencies of the rate of cavitation voids generation $\Gamma(r,t)$, normalized by $\Gamma_{norm} = \Gamma(\text{at}\,|P| = 30\text{ MPa}) \approx 5.9\cdot10^{25}$ $m^{-3}s^{-1}$. The curve 1 corresponds to t = 2 ns, 2 - 2.5ns, 3 - 3ns.

According to [1], the main reason of the breakdown developing in fluid in a nonuniform field at nanosecond times may be related to a rupture of the continuity of fluid induced by the electrostrictive forces in the inhomogeneous electric field. Thus, in the vicinity of a needle electrode, a region saturated by micropores is created. In the pores, the primary electrons are accelerated by the electric field to the energies exceeding the potential of ionization of a water molecule. The sources of primary electrons in the microcavities can include a background radiation, as well as the field emission from the surface of the cavities.

Figure 4 shows a schematic picture of the development of cavitation micropores arising in the vicinity of the electrode.

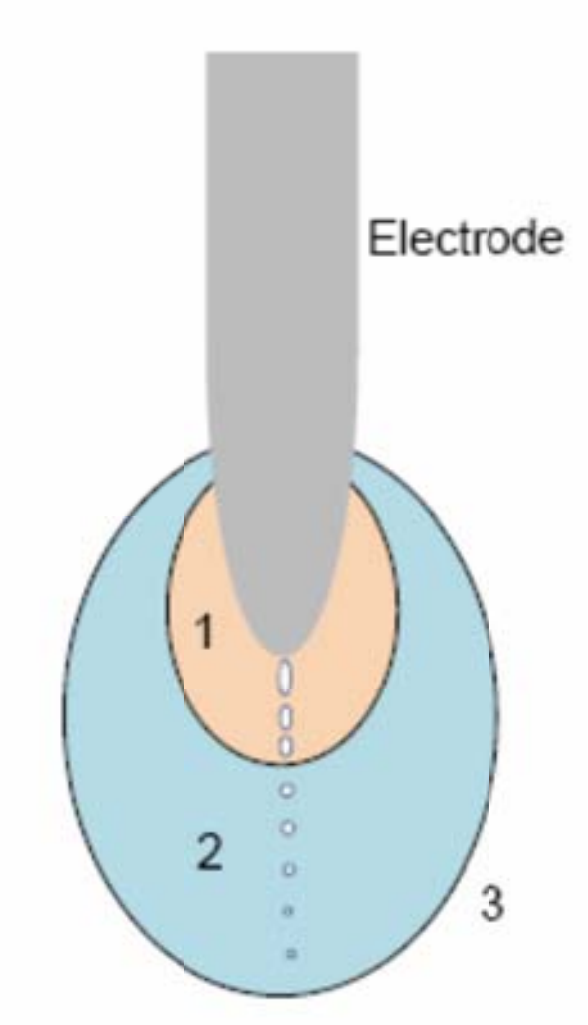

**Fig. 4.** Schematic picture of cavitation pore formation and growth in the vicinity of the electrode. In the region 1, where the electric field gradient is greatest, the occurring cavitation nanopores have enough time during the nanosecond voltage pulse to grow to a size at which an electron can gain enough energy for the excitation and the ionization of molecules of the liquid on the pore wall. In the region 2, the electrostrictive negative pressure reaches values at which the cavitation development becomes possible (which can be registered by the optical methods), but nanovoids appearing during the voltage pulse do not have enough time to grow to the size at which the potential difference across their borders becomes sufficient for the ionization or the excitation of water molecules. And, in the region 3 - the development of cavitation is impossible, since the spontaneously occurring nanovoids do not grow, because the value of the electrostrictive negative pressure is relatively small and cannot compete with the forces of surface tension.

## IV. Development of micropores in a dielectric liquid in inhomogeneous pulsed electric fields

Let us assume a spherical microcavity of the radius $R$ formed in water as a result of thermal fluctuations. Considering the discontinuity of the dielectric constant at the liquid-vacuum boundary, the surface forces exerted by the electric field on the surface of the pore per unit area are [15]:

$$F_{S,n} = \frac{\varepsilon_0}{2}\left(\alpha\left(\frac{E_{p,n}^2}{\varepsilon} + \varepsilon E_{p,t}^2\right) - (\varepsilon - 1)\left(E_{p,t}^2 + \frac{E_{p,n}^2}{\varepsilon}\right)\right) - \frac{2k_\sigma \sigma}{R} - p \ \text{[N/m}^2\text{]}. \quad (7)$$

Here, $E_{p,n}$, $E_{p,t}$ are the normal and tangential components of the electric field inside the pore. In (7) we took into account the surface tension of the liquid. Since the critical initial sizes of the cavitation pores are about a few nanometers, and the unperturbed electric field induced by the high voltage potential of the electrode is changing on a length scale of order of the electrode size (~ 10-100 microns, typical for experimental conditions in [3-8]). Therefore, the external unperturbed electric field $E_0$ in the vicinity of the pore can be considered as homogeneous with good accuracy

$$E_0/(dE_0/dr) >> R. \quad (8)$$

In this case, the electric field is uniform inside the ellipsoidal pore [13]. Wherein the expressions for the normal and tangential components of the electric field on the surface of the spherical pore are [13], [15]:

$$E_{p,n} = \frac{3\varepsilon}{1+2\varepsilon} E_0 \cos(\theta), \quad E_{p,t} = \frac{3\varepsilon}{1+2\varepsilon} E_0 \sin(\theta). \quad (9)$$

Here, $\theta$ is the azimuthal angle in a spherical coordinate system.

Since the objective of this work is to give the quantitative estimates and a qualitative picture of the development the cavitation micropores during the pulse strong nonuniform electric field, we will not consider changing the shape of the pores. Also, we assume that the force averaged over the surface acts on the surface of micropores. A form of the expanding micropore in a pulsed electric field will be discussed in Section V.

We obtain the mean pressure acting on the micropore surface, averaging (7) over the sphere surface, $4\pi R^2$:

$$P_{av} = (3/4)(\alpha - 1)\varepsilon_0 \varepsilon E^2 - 2k_\sigma \sigma / R - p, \quad (10)$$

Wherein, if $P_{av} < 0$, then the formed micropore collapses, but if $P_{av} > 0$, the micropore starts to expand.

Let us consider the problem of expanding the spherical pores under the influence of the electrostrictive ponderomotive forces and the surface tension. As was shown above, during the relatively short voltage pulse ($t_0 \sim 1-5$ ns), the fluid does not have enough time to be set in motion (see Fig.1 (b), 2 (b)), and, consequently, the hydrostatic pressure component in (10), $p$ **,** is much smaller than the electrostrictive. Therefore, we can neglect with the hydrostatic pressure. Next, we follow the analogy with the problem considered in the classic work of Rayleigh [27]. However, in contrast to [27], where the bubble was kept from collapsing by the gas pressure inside, in our case, the bubble is held by the electrostriction forces induced in the liquid.

The fluid velocity at the distance $r'$ from the center of an expanding nanopores follows from the continuity equation (1), at which $\rho$ is assumed to be constant, is:

$$u = \frac{R^2}{r'^2} U, \quad (11)$$

Here, $R$ is the radius of nanopores, $r' \geq R$ **;** $U$ is the rate of expansion. In this case, the kinetic energy of the fluid is:

$$W_K = \frac{1}{2} \int_{r=R}^{\infty} 4\pi\rho u^2 r'^2 dr' = 2\pi\rho U^2 R^4 \int_{r'=R}^{\infty} \frac{1}{r'^2} dr' = 2\pi\rho U^2 R^3 \quad (12)$$

Since the rate of expansion is $U = \frac{dR}{dt}$, the equation (12) can be rewritten as:

$$W_K = 2\pi\rho R^3\left(\frac{dR}{dt}\right)^2 . \tag{13}$$

The relevant work of the pressure forces (10) for the pores expansion is:

$$W_p = 4\pi\int_{R_0}^{R} P_{av} r^2 dr \tag{14}$$

Since the change in the kinetic energy of the fluid is equal to the work of the pressure forces, we obtain from (13) and (14):

$$\frac{d}{dt}\left(R^3\left(\frac{dR}{dt}\right)^2\right) = \frac{4}{\rho}R\frac{dR}{dt}\left(\frac{3}{8}R(\alpha-1)\varepsilon_0\varepsilon E^2 - k_\sigma\sigma\right) \tag{15}$$

In (15), we explicitly substituted $P_{av}$ with (10) and neglected the hydrostatic pressure component.

As in Part I, we assume that the electric field, from the spherical electrode to which the voltage is applied linearly, is increasing over time. Also, to simplify the formulas and the calculations presented below, we assume the parameter $k_\sigma(R) = \text{const} = 1$. This assumption is justified since we are interested in a significant expansion of the nanovoids compared to the size of the critical pores, i.e. $R >> R_{cr} \sim \delta$.

When substituting (4) into (15), we obtain:

$$\frac{d}{dt}\left(R^3\left(\frac{dR}{dt}\right)^2\right) = \frac{4\sigma}{\rho}R\frac{dR}{dt}\left(\frac{3\widetilde{P}_{\max}R}{4\sigma}\left(\frac{r_0}{r}\right)^4\left(\frac{t}{t_0}\right)^2 - 1\right)$$

$$\widetilde{P}_{\max} = 0.5\cdot(\alpha-1)\varepsilon\varepsilon_0\left(\frac{U_0}{r_0}\right)^2 \tag{16}$$

It is convenient to introduce a variable $x = (R/R_0)^{5/2}$, where $R_0$ is the initial size of the pores and $\tau = t/t_0$. In this case, the equation (16) can be rewritten as

$$\frac{d^2x}{d\tau^2} = 5\frac{\sigma\cdot t_0^2}{\rho R_0^3}x^{-1/5}\left(\frac{3\widetilde{P}_{\max}R_0}{4\sigma}\left(\frac{r_0}{r}\right)^4 x^{2/5}\tau^2 - 1\right). \tag{17}$$

It follows from (17) that at the distance from the center of the electrode, $r$, the micropore expands at $\tau \ge \tau_0(r) = \sqrt{\frac{3\widetilde{P}_{\max}R_0}{4\sigma}\left(\frac{r_0}{r}\right)^4 x^{2/5}}$. It is evident that since $\tau \le 1$, the region in which the pores can expand to is limited to: $r < r_0\left(\frac{3\widetilde{P}_{\max}R_0}{4\sigma}\right)^{1/4}$. Accordingly, at the time $\tau = \tau_0$, $x = 1$ and

$\frac{dx}{d\tau} = 0$.

All calculations presented below were conducted with the same parameters as in Section II: $r_0$ = 100 μm, $\varepsilon = 81$, $\alpha = 1.5$, $U_0$ = 54 keV, $t_0 = 3$ ns ), $\sigma = 0.072$ N/m, $R_0 = 2$ nm and correspond to the time moment $t = t_0 = 3$ ns.

Figure 5 shows the dependence of the maximum electrostrictive pressure $|P_E| = F(E,r) = \widetilde{P}_{\max}(r_0/r)^4$ and the pore radius $R$ on the distance from the electrode $\eta = r - r_0$. It is seen that, under these conditions, an area, in which, for example, the nanopores of the size 2 nm may be created and expanded, is slightly greater than 1 μm, whereas, the region, in which the tension generated by the ponderomotive force, is insufficient to cause cavitation at $\eta = r - r_0 > 14$ μm.

Figure 6 shows the dependence of the kinetic energy that an electron gets in the growing pore $\Delta\phi = 2E_{in}R$ depending on the distance from the pore to the electrode, where $E_{in} = \frac{3\varepsilon}{1+2\varepsilon}E_0 \approx 1.5E_0$ is the field inside the pore. It is seen that the electron kinetic energy can significantly exceed ~ 10 eV, i.e., the characteristic ionization potential of water molecules ($I_{H2O}$ ≈ 12.6 eV), in the region $\eta = r - r_0 < 0.43$ μm. Thus, the excitation (accompanied by UV and visible light) and the ionization starts in the vicinity of the electrode surface, in accordance with the observations of a nanosecond breakdown in the liquid (see, eg, [3-7]). It is apparent that the relative size of the area, in which $\Delta\phi$>10 eV, increases at a higher value $U_0$ or a smaller radius of the electrode $r_0$.

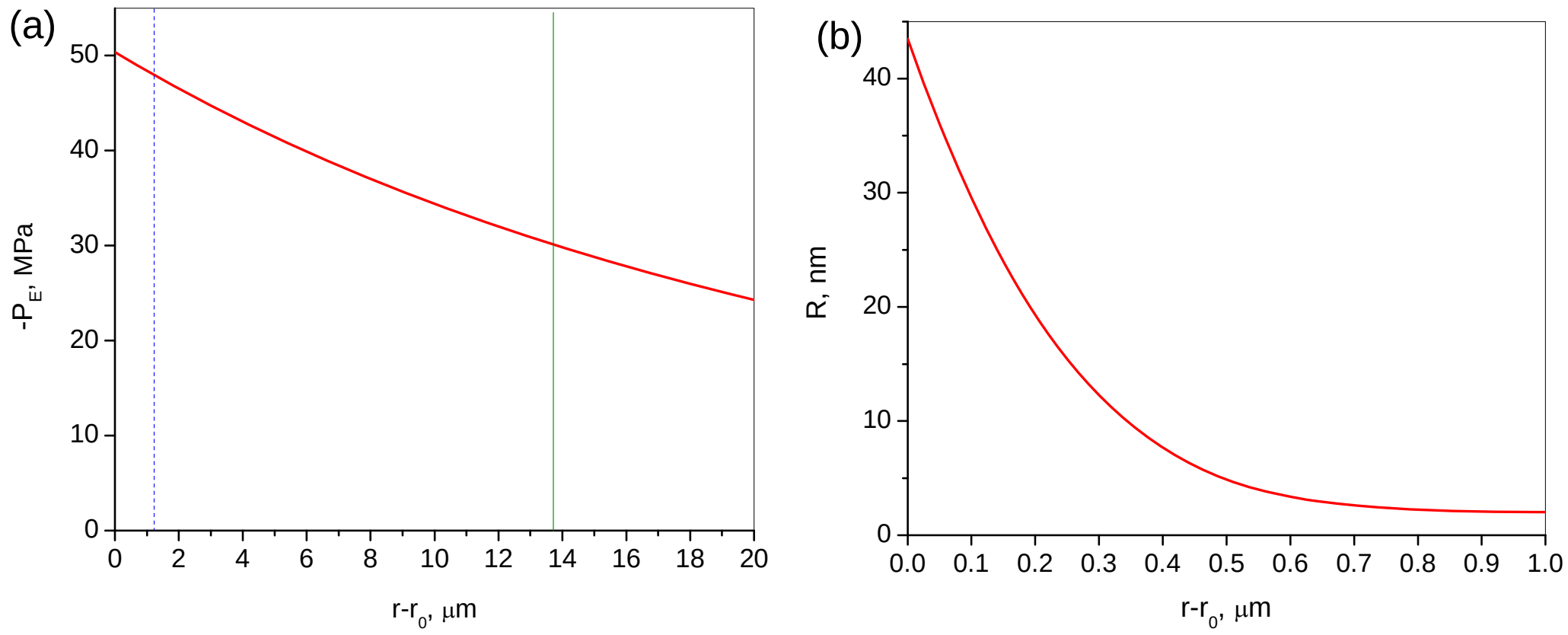

**Fig.5**. Dependencies of $P_E$ (a) and $R$ (b) on $\eta = r - r_0$ at the time moment $t = t_0 = 3$ ns. The area to the left of the dotted vertical line shows where the pores of the initial size 2nm are expanding, and to the right, it shows where they are collapsing. The area where the electrostriction tension (negative pressure) exceeds the cavitation threshold, $|P_E| > 30$ MPa [32] is to the left of the solid vertical line.

Figure 7 shows the dependence of the rate of micropores growth on time. It is seen that the rate of expansion is much less than the velocity of sound and much more than the characteristic velocity of the fluid under the action of ponderomotive forces. Therefore, the assumptions that during the voltage pulse a micropore center is not shifted and the discussed above modification of the Rayleigh model are applicable.

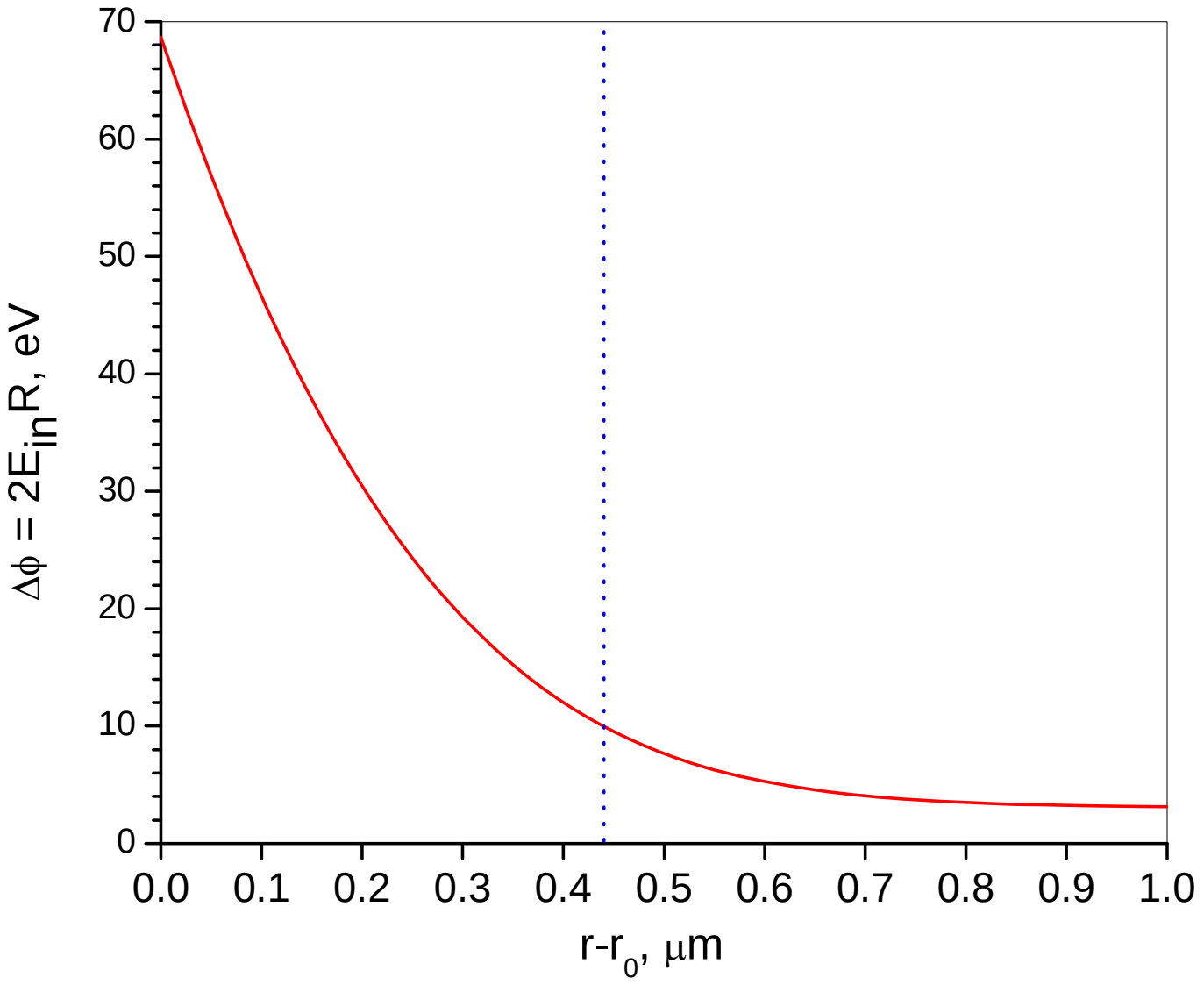

**Fig.6.** Parameter $\Delta\phi = 2E_{in}R$ dependence on $\eta = r - r_0$. The solid vertical line corresponds to $\Delta\phi = 10$ eV.

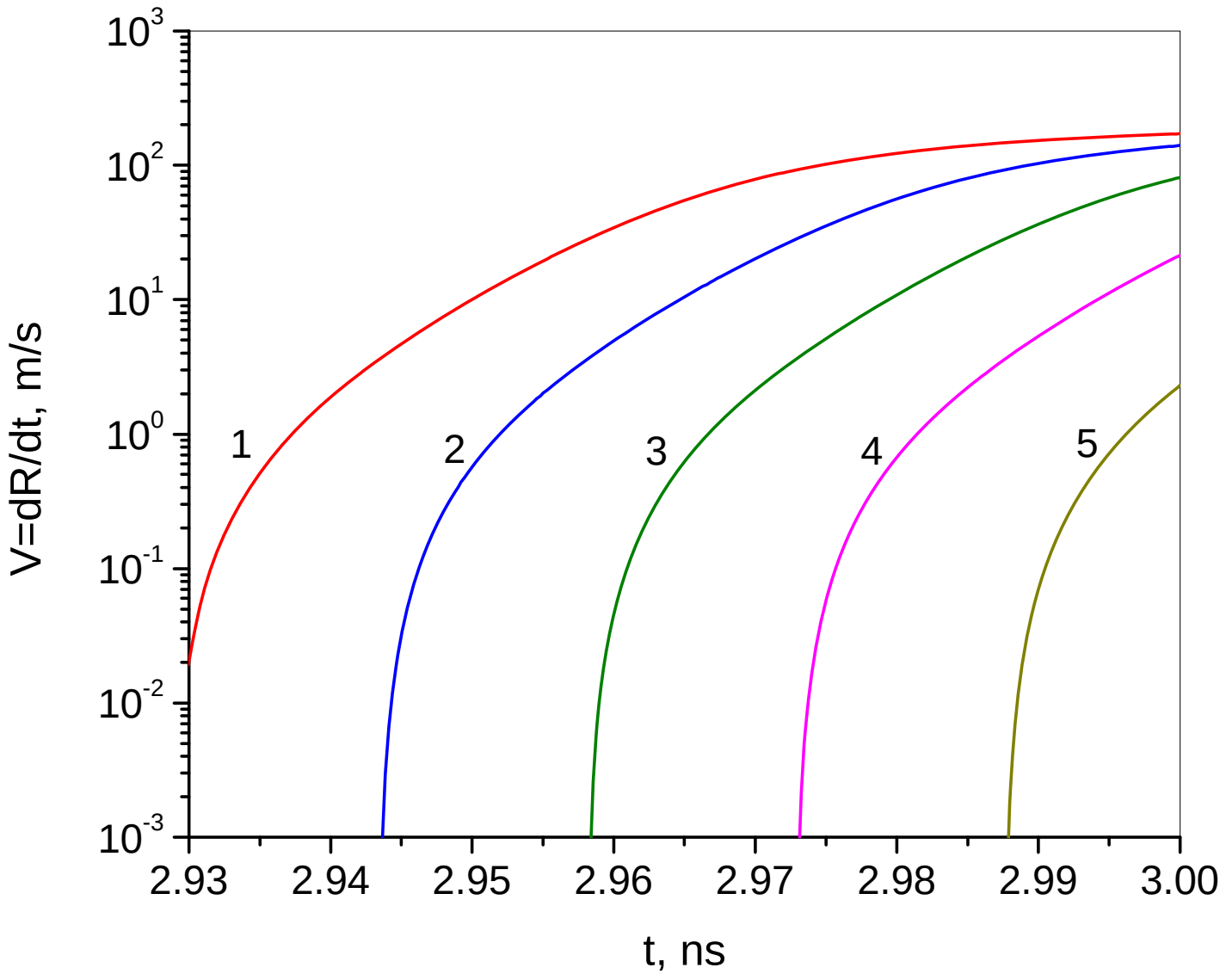

**Fig.7.** Dependencies of the rates of expansion of the pores on time at different $\eta = r - r_0$. Curve 1 corresponds to $\eta = 0$; 2 - $\eta = 0.25$; 3 - $\eta = 0.5$; 4 - $\eta = 0.75$ and 5 - $\eta = 1\,\mu\text{m}$.

## V. On the shape of the expanding micropores

We did not consider the fluid motion caused by the ponderomotive forces in the model of the nanopores expansion discussed above, since the rate of expansion (Fig.7) exceeds essentially the fluid velocity near the electrode (Fig. 1b, 2b). This statement is valid if the distortion of the electric field caused by the presence of the micropore is not taken into account. When considering the potential in the vicinity of the pores determined by an external potential and the dipole potential created by the polarization charges at the boundaries of the pores [13, 15], the expression for the square of the electric field near the spherical pore $E_{out}(r',\theta)$ in the assumption of a constant external electric field $E_0$ (induced by the applied voltage) is:

$$E_{out}^2(r',\theta) = E_0^2\left(1+\left(3\cos^2(\theta)+1\right)\left(\frac{(\varepsilon-1)}{2\varepsilon+1}\right)^2\frac{R^6}{r'^6}-\left(5\cos^2(\theta)-1\right)\frac{R^3}{r'^3}\frac{(\varepsilon-1)}{2\varepsilon+1}\right) \qquad (18)$$

Here, the angle $\theta = 0$ corresponds to the north pole of the pores (to the direction of the external unperturbed electric field), and $\theta = \pi/2$ relates to the equator (normal to the direction of the external unperturbed electric field). The dependencies $E^2/E_0^2$ on $r'/R$ at different angles $\theta$ are shown in Fig. 8:

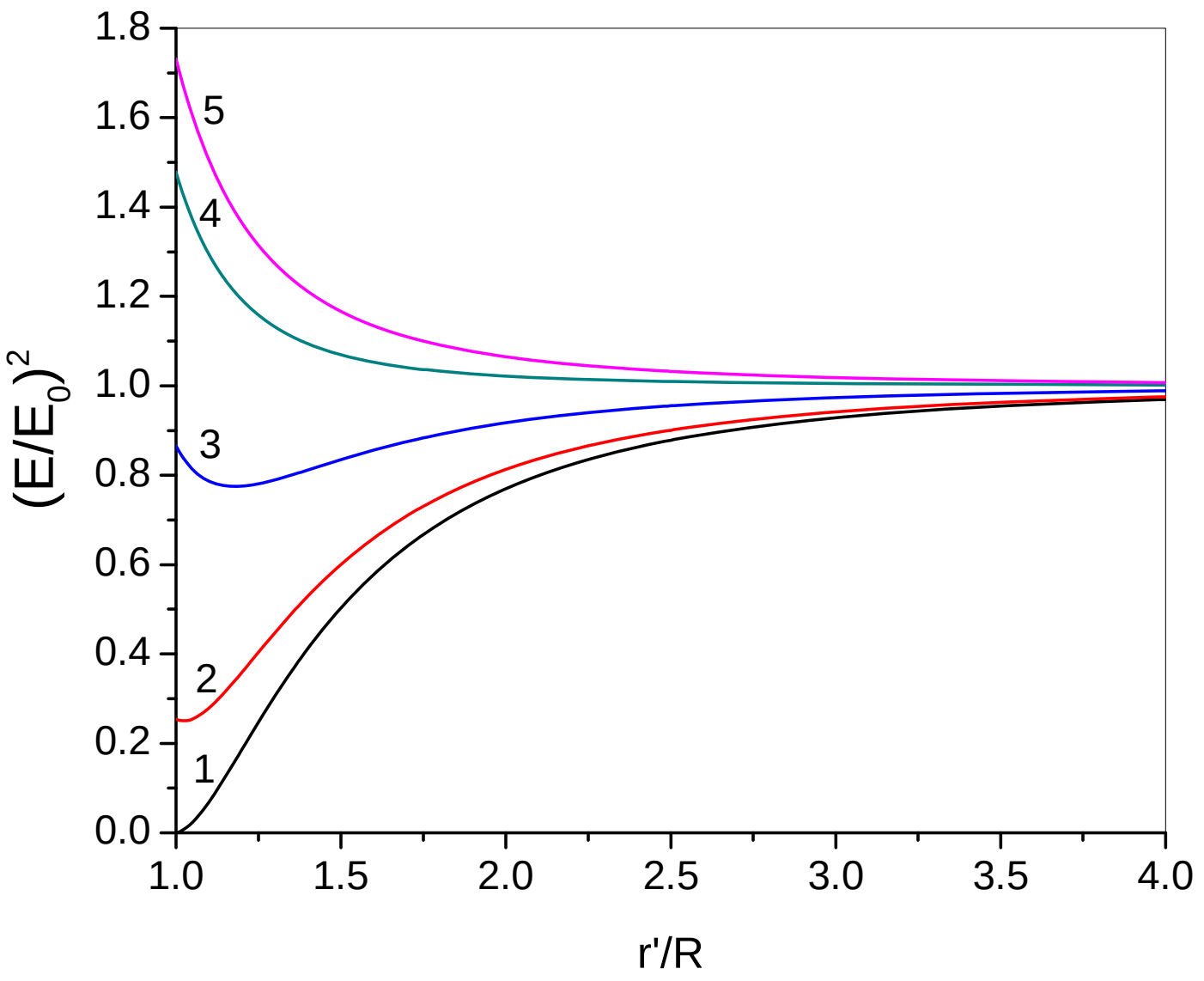

**Fig. 8**. Dependencies $E^2/E_0^2$ on $r'/R$ at different angles $\theta$. Line 1 coresponds to θ=0, 2 – θ= π/8, 3 – θ= π/4, 4 – θ= 3π/8, 5 – θ= π/2.

Taking into account (18) and (3) and neglecting the hydrostatic pressure and viscous friction, the volumetric electrostrictive force near the pore in the equation of fluid motion (1) has the form:

$$F(r',\theta) \approx \frac{3}{2}\alpha\varepsilon\varepsilon_0 E_0^2\frac{(\varepsilon-1)}{2\varepsilon+1}\frac{R^3}{r'^4}\left(\left(5\cos^2(\theta)-1\right)-2\left(3\cos^2(\theta)+1\right)\frac{R^3}{r'^3}\frac{(\varepsilon-1)}{2\varepsilon+1}\right). \qquad (19)$$

The dependencies on the volumetric force $F(r',\theta)$, referred to

$F_0 = \frac{3}{2}\alpha\varepsilon\varepsilon_0\frac{(\varepsilon-1)}{2\varepsilon+1}\frac{E_0^2}{R}\left(1+2\frac{(\varepsilon-1)}{2\varepsilon+1}\right)$, are shown in Fig.9.

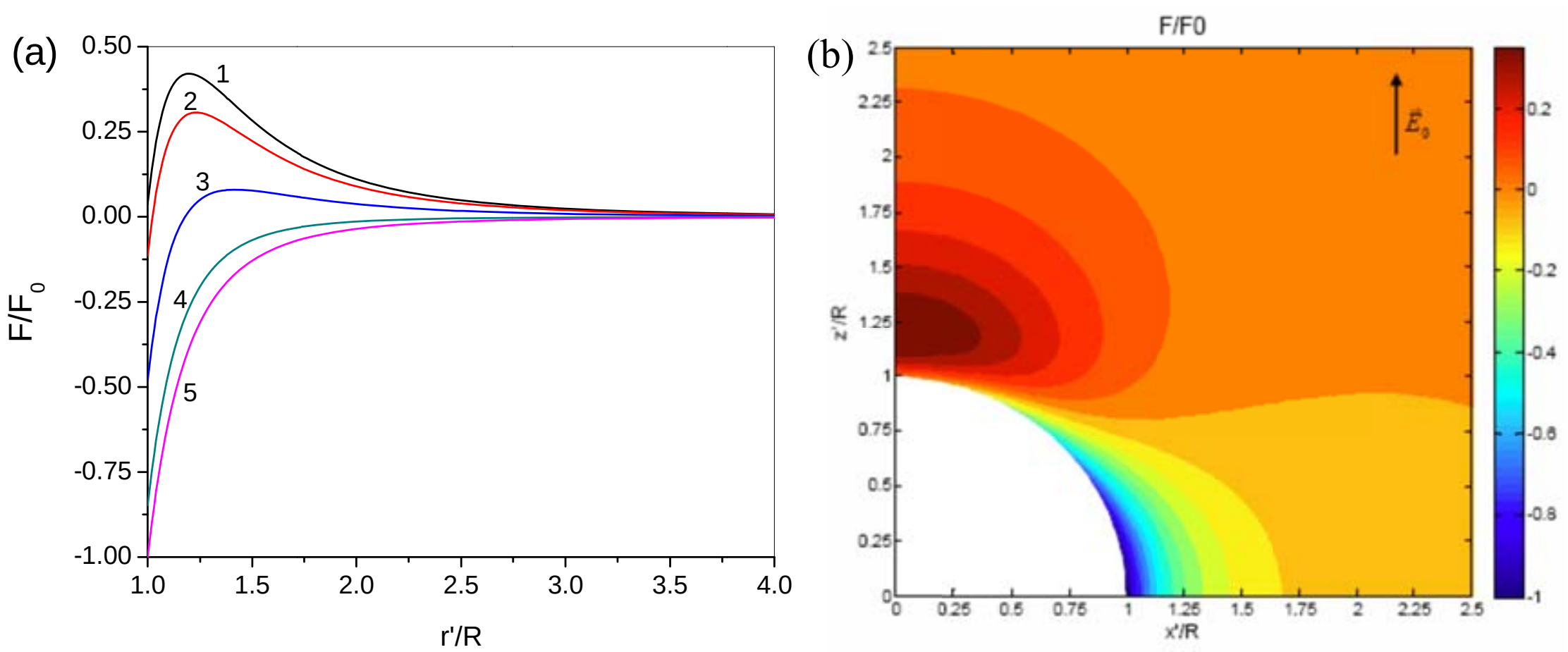

**Fig. 9.** Normalized volumetric force $F(r',\theta)$ in the vicinity of the pores at different angles $\theta$ . a) – radial dependencies. Line 1 coresponds to θ=0, 2 – θ= π/8, 3 – θ= π/4, 4 – θ= 3π/8, 5 – θ= π/2; b) Two-dimensional contour plot

The fluid moves toward the pore in the vicinity of the equator, due to the volumetric electrostrictive force directed toward the center of the pore (negative direction). However, at the pole, it moves away from the pore, because the electrostrictive force is positive. Therefore, the micropore will be stretched along the electric field, since its expansion in the equatorial plane will be inhibited by the fluid counterflow at the equator (Fig. 10).

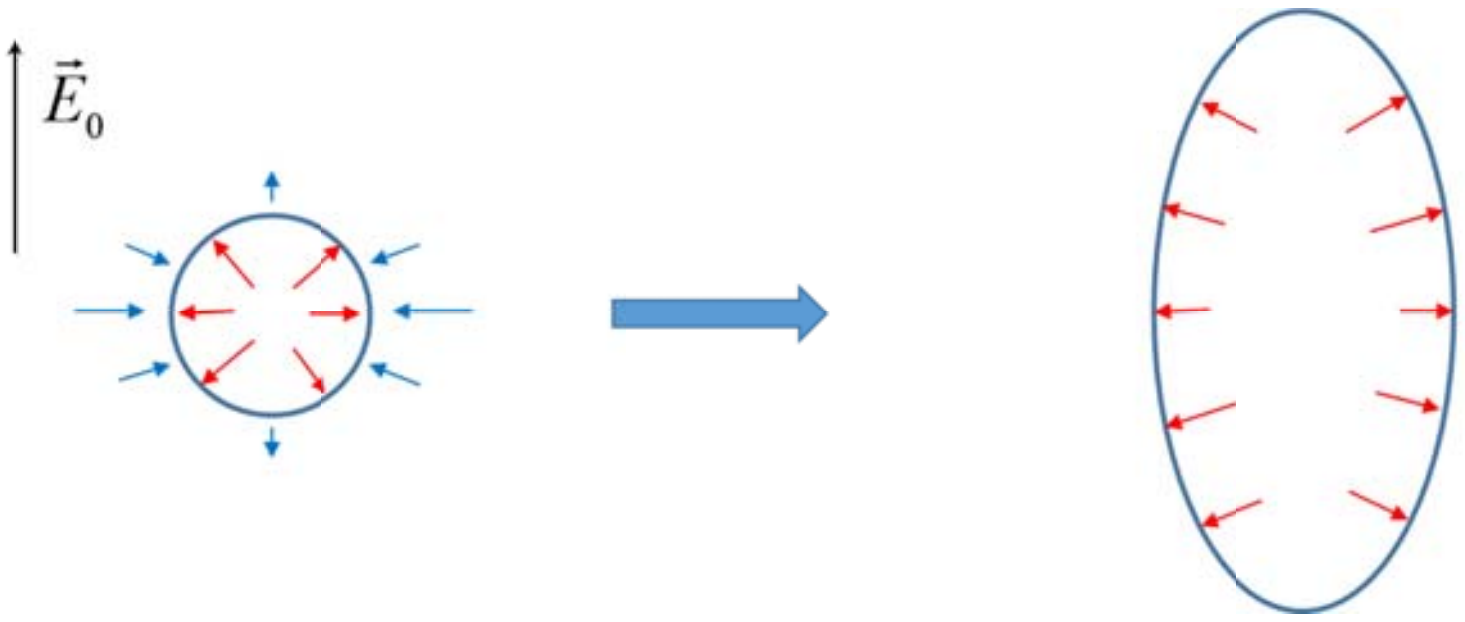

**Fig. 10.** Qualitative picture of the micropores form formation. $\vec{E}_0$ is the unperturbed external field produced by the voltage applied to the electrode. The red arrows indicate the "electrostatic" pressure inside the pores, the blue arrows demonstrate the directions of the fluid motion in the outer region of the pore induced by the electrostrictive volumetric force.

It should be noted that the above considerations are mostly qualitative and valid only for the fluid outside the pore assumed as an elastic medium, i.e., for a very small deformation of the micropore and fluid in its vicinity. In order to determine the exact shape of the expanding pore we need to consider in the vicinity of the pore following factors, such as: boundary conditions for the ponderomotive forces, tension forces, the hydrostatic pressure, and, finally, the appearance of new microscopic discontinuities (nanopores) in the fluid in the region of large tensile stresses.

## Conclusions

1. The equations describing the nanopore expansion in the presence of a strong inhomogeneous pulsed electric field are obtained, and the corresponding approaches are formulated.
2. It is shown that the three characteristic regions appear in a liquid dielectric in the vicinity of

the electrode. In the first region, where the electric field gradient is greatest, the occurring cavitation nanopores have enough time during the nanosecond voltage pulse to grow to a size at which an electron can gain enough energy for the excitation and ionization of the liquid molecules on the pore wall. In the second region, the electrostrictive negative pressure reaches values at which cavitation development becomes possible (and can be recorded by the optical methods). However, the nanovoids appearing during the voltage pulse do not have enough time to grow to the size at which the potential difference across their borders becomes sufficient for the ionization or excitation of water molecules. And, in the third region, the development of cavitation is impossible, since the spontaneously occurring nanovoids do not grow, because the value of the electrostrictive negative pressure is relatively small and cannot compete with the forces of surface tension.

3. The physical processes affecting the shape of the expanding nanopores are considered. It is shown that a perturbation of the electric field caused by a micropore near its equator (in the plane perpendicular to the direction of the unperturbed external electric field) produces the fluid flow directed towards the pore, and in the vicinity of poles, away from the pore. As a consequence, a micropore will extend along the external electric field.